\documentclass[aps,prd,onecolumn,showpacs,superscriptaddress,preprintnumbers,floatfix,reprint,nofootinbib]{revtex4-1}
\usepackage{graphicx}
\usepackage{amsmath}
\usepackage{bm}
\usepackage{yhmath}
\usepackage[colorlinks,citecolor=blue,urlcolor=blue,linkcolor=blue]{hyperref}
\usepackage{subfigure}
\usepackage{color}
\usepackage{cases}
\usepackage{subfigure}
\usepackage{dcolumn,booktabs,bm}
\usepackage{slashed}
\usepackage{amsfonts,amssymb,stmaryrd,latexsym,amsmath}
\usepackage{textcomp}
\usepackage{multirow}
\usepackage{cancel}
\usepackage{diagbox}
\usepackage{array}

\renewcommand{\arraystretch}{1.8}
\allowdisplaybreaks

\begin{document}

\title{Mass spectrum and strong decays of tetraquark $\bar c\bar s qq$ states}
\author{Guang-Juan Wang}\email{wgj@pku.edu.cn}
 \affiliation{Center of High Energy Physics, Peking University, Beijing 100871, China}

\author{Lu Meng}\email{lmeng@pku.edu.cn}
\affiliation{Center of High Energy Physics, Peking University,
Beijing 100871, China}

\author{Li-Ye Xiao}\email{lyxiao@ustb.edu.cn}
\affiliation{School of Mathematics and Physics, University of
Science and Technology Beijing, Beijing 100083, China}
\author{Makoto Oka}\email{oka@post.j-parc.jp}
\affiliation{Advanced Science Research Center, Japan Atomic Energy Agency, Tokai, Ibaraki, 319-1195, Japan}
\affiliation{Nishina Center for Accelerator-Based Science, RIKEN, Wako 351-0198, Japan}

\author{Shi-Lin Zhu}\email{zhusl@pku.edu.cn}
\affiliation{School of Physics and State Key Laboratory of Nuclear
Physics and Technology, Peking University, Beijing 100871, China}

\begin{abstract}
We systematically study the mass spectrum and strong decays of the
S-wave $\bar c\bar s q q$ states in the compact tetraquark scenario
with the quark model. The key ingredients of the model are the
Coulomb, the linear confinement, and the hyperfine interactions. The
hyperfine potential leads to the mixing between different color
configurations, as well as the large mass splitting between the two
ground states with $I(J^P)=0(0^+)$ and $I(J^P)=1(0^+)$. We calculate
their strong decay amplitudes into the $\bar D^{(*)}K^{(*)}$
channels with the wave functions from the mass spectrum calculation
and the quark interchange method. We examine the interpretation of
the recently observed $X_0(2900)$ as a tetraquark state. The mass
and decay width of the $I(J^P)=1(0^+)$ state are $M=2941$ MeV and
$\Gamma_X=26.6$ MeV, respectively, which indicates that it might be
a good candidate for the $X_0(2900)$. Meanwhile, we also obtain an
isospin partner state $I(J^P)=0(0^+)$ with $M=2649$ MeV and
$\Gamma_{X\rightarrow \bar D K}=48.1$ MeV, respectively. Future
experimental search for $X(2649)$ will be very helpful.

\end{abstract}
\maketitle

\section{introduction}\label{intro}

Recently, the LHCb collaboration reported an enhancement on the $D^-
K^+$ invariant mass distribution in the process $B^+\rightarrow D^+
D^- K^+$, which is parameterized as two Breit-Wigner
resonances~\cite{Joh:2020,Aaij:2020hon, Aaij:2020ypa}:
\begin{eqnarray}
&&X_{0}:\nonumber\\
 &&M=2.866\pm0.007\pm 0.002~\text{GeV},\,\Gamma=57\pm 12\pm 4~\text{MeV}, \nonumber\\
&&X_{1}: \,\,\ \nonumber\\
&&M=2.904 \pm 0.005 \pm 0.001~\text{GeV},    \,\Gamma=110 \pm 11 \pm  4~\text{MeV}, \nonumber
\end{eqnarray}
with the $J^P=0^+$ and $J^P=1^-$, respectively. Since they were
observed in the $D^- K^+$ channel, the minimal quark contents are
$\bar c \bar s d u$. In 2016, another open flavor exotic state
$X(5586)$ with the quark contents $su\bar b\bar d$ (or $sd\bar b\bar
u$) was reported by the D0 collaboration. However, it was not
confirmed by the LHCb~\cite{Aaij:2016iev},
CMS~\cite{Sirunyan:2017ofq}, CDF~\cite{Aaltonen:2017voc} and
ATLAS~\cite{Aaboud:2018hgx} collaborations. Thus, $X(2900)$ states
might be the first open flavor tetraquark states in experiment,
which deserve a refined investigation on its existence and the inner
dynamics. So far, their inner structures and properties are still
not clear. One explanation is that the enhancement might be solely
from the rescattering effects of the $\bar D^*K^*\rightarrow \bar
DK$ or $\bar D_1K^{(*)}\rightarrow \bar DK$
channels~\cite{Liu:2020orv,Burns:2020epm,Chen:2020eyu}. Another
explanation is the genuine
resonance~\cite{Molina:2010tx,Hu:2020mxp,Liu:2020nil,He:2020btl,Huang:2020ptc,Xue:2020vtq,Molina:2020hde,Agaev:2019wkk,Agaev:2020nrc,Karliner:2020vsi,Wang:2020xyc,Zhang:2020oze,He:2020jna,Chen:2020aos,Mutuk:2020igv,Albuquerque:2020ugi},
either the loosely bound molecular states or the compact tetraquark
states.

In the hadronic molecular scheme, the $X_0(2900)$ is explained as a
$\bar D^* K^*$ molecule with $J ^P= 0^+$ in
Refs.~\cite{Hu:2020mxp,Liu:2020nil,He:2020btl,Huang:2020ptc,Xue:2020vtq,Molina:2020hde,Agaev:2019wkk,Chen:2020aos,Mutuk:2020igv}.
The molecular assignment of $X_1(2900)$ is ruled out in
Refs.~\cite{Liu:2020nil,Huang:2020ptc} and it is explained as a
virtual state in Ref.~\cite{He:2020btl}. Among these work, the
isospin of the $X_0(2900)$ is determined as $I=0$ in Refs.
\cite{Hu:2020mxp,Liu:2020nil,Xue:2020vtq,Molina:2020hde}, while in
Ref.~\cite{He:2020btl}, its isospin is found to be $I=1$. The
isospin of $X_0(2900)$ is not presented in the QCD sum rule
calculation with the meson-meson interpolating
current~\cite{Chen:2020aos,Agaev:2020nrc}.

Within the tetraquark scheme, the $X_0(2900)$ is accommodated as the
tetraquark state with
$J^P=0^+$~\cite{Karliner:2020vsi,Wang:2020xyc,Zhang:2020oze,He:2020jna},
while the quark model study with dynamical
calculation~\cite{Lu:2020qmp} does not favor it in view of the mass
differences between the experimental mass of the $X_0(2900)$ and the
predicted masses $(2675, 3065, 3152, 3396)$ MeV. The $X_1(2900)$ is
interpreted as the tetraquark state in
Refs.~\cite{Chen:2020aos,Mutuk:2020igv}.

The nature of the $X(2900)$ states is still controversial. Most of
the investigations mentioned above focus on the mass spectroscopy of
$X(2900)$, which is a major platform to probe the dynamics of the
multi-quark system. However, compared with the conventional baryons
and mesons, the tetraquark states are more complicated. For example,
the tetraquark states contain much richer color configurations,
i.e., $\bar 3_c-3_c$ and $6_c-\bar 6_c$ components
~\cite{Wang:2019rdo}. Therefore, the quark models that succeed in
the conventional mesons $(\bar q q)$ and baryons $qqq$, might have
large uncertainties in the multi-quark systems. The other properties
such as production and decay patterns become essential, which are
very sensitive to the inner structures of the multiquark system. The
authors in Ref.~\cite{Burns:2020xne} discussed the productions and
the decays of the $X(2900)$ in different physical scenarios,
including the triangle diagrams, the molecules and the tetraquarks.

In this paper, we study the four-quark state $\bar c\bar sq q$ in
the tetraquark scenario and examine whether the $X_0(2900)$ can be
accommodated as a compact tetraquark. We first extend our previous
work~\cite{Wang:2019rdo} to study the open-flavor system $\bar c
\bar s q q$ spectrum within quark model considering the mixing
effect of the two color configurations. Then, we calculate the decay
widths of the $\bar c\bar s qq$ system into the $\bar D^{(*)}
K^{(*)}$ channels within the quark interchange
model~\cite{Wong:2001td,Barnes:1991em,Swanson:1992ec,Barnes:1999hs,Barnes:2000hu,Wang:2018pwi}.
In this model, the quark-quark interactions are described by thes
same quark model in the mass spectrum calculation.

The paper is arranged as follows. In Sec.~\ref{sec1}, we introduce
the Hamiltonian for the $\bar c \bar s q q$ system. With the quark
model, we calculate the mass spectrum of the S-wave tetraquark
states and present the results in Sec.~\ref{sec2}. With the same
quark-quark interaction, we investigate their strong decays into the
$\bar D^{(*)} K^{(*)}$ channel using the quark interchange model.
The details are elaborated in Sec.~\ref{sec3}. We give a summary in
Sec.~\ref{sec4}.

\section{Hamiltonian}\label{sec1}

In a tetraquark state, the four-quark Hamiltonian reads,
\begin{eqnarray}
H & =&
\sum_{i=1}^{4}\frac{p_{j}^{2}}{2m_{j}}-T_G+\sum^4_{i<j=1}V_{ij}+\sum_{j}m_{j},
\end{eqnarray}
with $p_j$ and $m_j$ denote the momentum and mass of the inner quark
with index $j$, respectively. $T_G$ is the total kinematic energy of
the system and vanishes in the center-of-mass frame. The $V_{ij}$ is
the quark-quark interaction between the quark pair $(ij)$. In this
work, we use the non-relativistic quark model proposed in
Ref.~\cite{SilvestreBrac:1996bg} to describe the quark-quark
interaction,
\begin{eqnarray}\label{qm}
&&V_{ij}(r_{ij})=-\frac{3}{16}\sum_{i<j}\mathbf{\lambda}_{i}\mathbf{\lambda}_{j}\Big(-\frac{\kappa(1-\text{exp}(-r_{ij}/r_{c}))}{r_{ij}}+b r_{ij}^{p}\nonumber \\
&&-\Lambda+\frac{8\pi}{3m_{i}m_{j}}\kappa'(1-\text{exp}(-r_{ij}/r_{c}))\frac{\text{exp}(-r_{ij}^{2}/r_{0}^{2})}{\pi^{3/2}r_{0}^{3}}\mathbf{s}_{i}\cdot\mathbf{s}_{j}\Big),\nonumber \\
\end{eqnarray}
with
\begin{eqnarray}
&&r_0=A(\frac{2m_im_j}{m_i+m_j})^{-B}, \nonumber
\end{eqnarray}
\begin{eqnarray}
&&A=1.6553~\text{GeV}^{B-1},~B=0.2204,\nonumber
\end{eqnarray}
\begin{eqnarray}
r_{c}=0,~p=1,~\Lambda=0.8321\text{GeV},\nonumber
\end{eqnarray}
\begin{eqnarray}
\kappa=0.5069,~b=0.1653~\text{GeV}^{2},~\kappa'=1.8609,\nonumber
\end{eqnarray}
\begin{eqnarray}
&&m_{c}=1.836~\text{GeV}, m_{q}=0.315~\text{GeV},
m_{s}=0.577~\text{GeV},\nonumber\\ \label{para}
\end{eqnarray}
where $r_{ij}$ is the radius between the $i$th and $j$th quarks. The $\mathbf{\lambda}_{i}$ ($-\mathbf{\lambda}_i^T$) is the color generator for the quark (antiquark). The first two terms are the central parts of the model, the Coulomb and linear confinement potentials. The last term in the $V_{ij}$ is the hyperfine interaction with the smearing effect parameterized by $r_0$, which is related to the reduced mass of the interacting quarks. The $\mathbf s _i$ is the spin operator of the $i$th quark. The hyperfine potential contains all the flavor information and is expected to play an important role for the mass spectrum. The values of the parameters are determined by fitting the mass spectra of the mesons in Table~\ref{par}.
\begin{table*}
 \renewcommand\arraystretch{1.5}
 \caption{The mass spectra of the mesons (in units of MeV) in the quark model. The experimental
 results are taken from Ref.~\cite{pdg}. }\label{par}
 \centering
 \setlength{\tabcolsep}{1.8mm}
\begin{tabular}{c|cccccccccc}
\toprule[1pt]
 & $D_{s}$ & $D_{s}^{*}$ & $D$ & $D^{*}$ & $K$ & $K^{*}$ & $\pi$ & $\rho$\tabularnewline
\hline Exp.~\cite{pdg} & $1968.3$ & $2112.2$ & $1869.7$ & $2010.3$ &
$493.7$ & $891.7$ & $139.6$ & $775.3$\tabularnewline \hline Theo. &
$1963.0$ & $2102.4$ & $1862.8$ & $2016.3$ & $491.3$ & $903.7$ &
$138.5$ & $770.1$\tabularnewline \bottomrule[1pt]
\end{tabular}
\end{table*}

\section{mass spectrum}\label{sec2}
\subsection{Wave function}\label{sec2:1}
\begin{figure*}[htbp]
\centering
\includegraphics[width=0.6\textwidth]{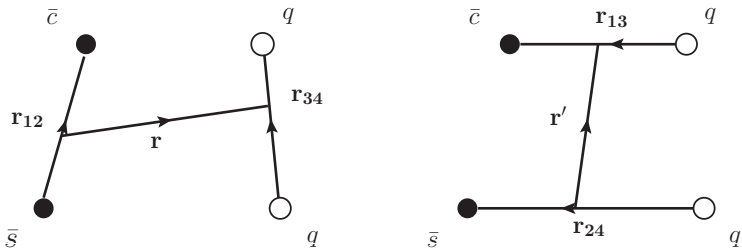}
\caption{The Jacobi coordinates in the tetraquark state $\bar c\bar
s qq$. } \label{jac}
\end{figure*}

The spatial wave function of the four-body system can be described
by three independent Jacobi coordinates. For the $\bar c \bar s q q$
state, there are two sets of Jacobi coordinates as presented in
Fig.~\ref{jac}. In general, if we use one set of Jacobi coordinates
with complete spatial bases and the complete color configurations,
we can describe the four-quark system. In this work, we use the
Jacobi coordinate in Fig.~\ref{jac} (a) to expand the wave function
of the tetraquark state.
It reads
\begin{eqnarray}\label{wavefunction}
&&\Psi^{II_z}_{JJ_z}=\sum_{\alpha}A_{\alpha}\psi_{\alpha}(\mathbf{ r_{12}},\mathbf{ r_{34}},\mathbf{r}),\nonumber \\
&&\psi_{\alpha}=\phi_{n_al_a}({r}_{12},\beta_a)\phi_{n_bl_b}({r}_{34},\beta_b)\phi_{n_{ab}l_{ab}}({r},\beta)\nonumber\\
&&\otimes \left[[Y_{l_a}(\hat {\mathbf{r_{12}}})Y_{l_b}(\hat {\mathbf{r_{34}}})]_{\ell}Y_{l_{ab}}(\hat {\mathbf{r}})]_L\otimes[\chi_{s_{a}}(12)\chi_{s_{b}}(34)]_S\right]_{JJ_z} \nonumber\\
&&\otimes [\chi_c(12)\chi_c(34)]_{1_c}
\otimes[\xi_f(12)\xi_f(34)]_{II_z},
\end{eqnarray}
where the $(1,2,3,4)$ denote the four quarks $(\bar c, \bar s, q,q
)$, respectively. $J$ $(J_z)$ and $I$ $(I_z)$ are the total angular
momentum and the isospin (the third component) of the tetraquark
state. The $l_{a}$, $l_b$, and $l_{ab}$ represent the orbital
angular momentum within the $(\bar c\bar s)$ cluster, the $(q q)$
cluster and that between these two clusters, respectively. The $l_a$
and $l_b$ couples into the orbital angular momentum $\ell$. Then
$\ell$ couples with the $l_{ab}$ into the total orbital angular
momentum $L$. The $s_a$ and $s_b$ are the spin of the $(\bar c\bar
s)$ and $(qq)$ clusters. They form the total spin of the tetraquark
$S$. The $Y_{lm}$ is the spherical harmonics function. The sum of
the script $\alpha$ represents the superposition of the four-body
bases that satisfy the quantum number ($JJ_z,II_z$). $A_{\alpha}$ is
the expanding coefficient for the corresponding basis. The
$\chi_{c}$, $\chi_s$, and $\xi$ are the wave functions in the color,
spin and flavor space, respectively. The subscript $1_c$ denotes
that the four quarks form a color singlet state. The $\phi$ is the
radial wave function in spatial space and reads
\begin{eqnarray}
\phi_{n_a l_a}(\mathbf{r}_{12},\beta_a)&=&\Big\{
\frac{2^{l_a+2}(2\nu_{n_a})^{l_a+3/2}}{\sqrt \pi (2l_a+1)!!}\Big
\}^{1/2}r_{12}^{l_{a}} e^{-\nu_{n_a} r_{12}^{2}},\nonumber
\end{eqnarray}
where $\nu_{n_a}$ is related to the oscillating parameter $\beta_a$
as follows,
\begin{eqnarray}
\nu_{n_a}=\frac{n_a \beta^2_a}{2}, \,\,\,
(n_a=1,2,..,n^{\text{max}}_a),
\end{eqnarray}
the $n_a$ is the radial quantum number. $n^{\text{max}}_a$ is the
number of the expanding bases along $r_a$. The spatial wave
functions $\phi_{n_b l_b}(\mathbf{r}_{34},\beta_b)$ and
$\phi_{n_{ab} l_{ab}}(\mathbf{r},\beta_{ab})$ have the similar
forms. In this work, we concentrate on the S-wave tetraquark state,
which should be a superpositions of the states
$|l_a=l_b=l=0\rangle$, $|l_a=l_b=1,l=0\rangle$,
$|l_a=2,l_b=l=0\rangle$ etc.. In conventional hadrons, the mass gap
between the ground state and its first orbital excitation is about
hundreds of MeV. The same pattern is expected in the compact
multi-quark states. In addition, the high orbital excitations couple
with the ground states through the tensor and spin-orbital
potentials, which can be treated as perturbative compared with the
Coulomb and linear confinement potentials. Thus, we only expand the
tetraquark states by the ground spatial bases.

In the state $\bar c\bar sqq$, the color-spin-flavor wave
configuration is constrained by the Fermi-statistic and the possible
wave functions are listed in Table~\ref{csfwavefunction}.
\begin{table*}
 \renewcommand\arraystretch{1.8}
 \caption{The color-spin-flavor wave functions of the S-wave tetraquark states $\bar c \bar s q_1 q_2$ with different $J^P$ quantum numbers. The subscripts and superscripts denote the color representation and the spin, respectively. The symbol $\{q_1q_2\}$ and $[q_1q_2]$ represent that the two light quarks are symmetric and antisymmetric in the flavor space, respectively. The $(\beta_{a},\beta_{b},\beta_{ab})$ are the oscillating parameters in the spatial wave functions of the antidiquark $\bar c\bar s$, the diquark $q_1q_2$ and that between the two clusters. }\label{csfwavefunction}
 \centering
 \setlength{\tabcolsep}{2.3mm}
\begin{tabular}{c|c|c|c}
\toprule[1pt] \multicolumn{2}{c|}{$I=1$} &
\multicolumn{2}{c}{$I=0$}\tabularnewline \midrule[1pt]
\multirow{2}{*}{$J^{P}=0^{+}$} & $\Psi_{1}=\left[(\bar c\bar
s)_{{3}_{c}}^{1}\{ {q}_1{q}_2\}_{\bar
3_{c}}^{1}\right]_{1_{c}}^{0}\psi(\beta_{a},\beta_{b},\beta_{ab})$ &
\multirow{2}{*}{$J^{P}=0^{+}$} & $\Psi_{1}=\left[(\bar c\bar
s)_{{3}_{c}}^{0}[{q}_1{q}_2]_{\bar
3_{c}}^{0}\right]_{1_{c}}^{0}\psi(\beta_{a},\beta_{b},\beta_{ab})$\tabularnewline
 & $\Psi_{2}=\left[(\bar c\bar s)_{\bar6_{c}}^{0}\{{q}_1{q}_2\}_{{6}_{c}}^{0}\right]_{1_{c}}^{0}\psi(\gamma_{a},\gamma_{b},\gamma_{ab})$ & & $\Psi_{2}=\left[(\bar c\bar s)_{\bar 6_{c}}^{1}[{q}_1{q}_2]_{{6}_{c}}^{1}\right]_{1_{c}}^{0}\psi(\gamma_{a},\gamma_{b},\gamma_{ab})$\tabularnewline
\hline \multirow{3}{*}{$J^{P}=1^{+}$} & $\Psi_{1}=\left[(\bar c\bar
s)_{{3}_{c}}^{1}\{ {q}_1{q}_2\}_{ \bar
3_{c}}^{1}\right]_{1_{c}}^{1}\psi(\beta_{a},\beta_{b},\beta_{ab})$ &
\multirow{3}{*}{$J^{P}=1^{+}$} & $\Psi_{1}=\left[(\bar c\bar
s)_{{3}_{c}}^{1}[{q}_1{q}_2]_{\bar
3_{c}}^{0}\right]_{1_{c}}^{1}\psi(\beta_{a},\beta_{b},\beta_{ab})$\tabularnewline
 & $\Psi_{2}=\left[(\bar c\bar s)_{{3}_{c}}^{0}\{ {q}_1{q}_2\}_{\bar 3_{c}}^{1}\right]_{1_{c}}^{1}\psi(\lambda_{a},\lambda_{b},\lambda_{ab})$ & & $\Psi_{2}=\left[(\bar c\bar s)_{\bar 6_{c}}^{1}[{q}_1{q}_2]_{ {6}_{c}}^{1}\right]_{1_{c}}^{1}\psi(\lambda_{a},\lambda_{b},\lambda_{ab})$\tabularnewline
 & $\Psi_{3}=\left[(\bar c\bar s)_{\bar 6_{c}}^{1}\{ {q}_1{q}_2\}_{{6}_{c}}^{0}\right]_{1_{c}}^{1}\psi(\gamma_{a},\gamma_{b},\gamma_{ab})$ & & $\Psi_{3}=\left[(\bar c\bar s)_{\bar 6_{c}}^{0}[{q}_1{q}_2]_{{6}_{c}}^{1}\right]_{1_{c}}^{1}\psi(\gamma_{a},\gamma_{b},\gamma_{ab})$\tabularnewline
\hline $J^{P}=2^{+}$ & $\Psi_{1}=\left[(\bar c\bar
s)_{{3}_{c}}^{1}\{ {q}_1{q}_2\}_{\bar
3_{c}}^{1}\right]_{1_{c}}^{2}\psi(\beta_{a},\beta_{b},\beta_{ab})$ &
$J^{P}=2^{+}$ & $\Psi_{1}=\left[(\bar c\bar s)_{\bar
6_{c}}^{1}[{q}_1{q}_2]_{{6}_{c}}^{1}\right]_{1_{c}}^{2}\psi(\beta_{a},\beta_{b},\beta_{ab})$\tabularnewline

\bottomrule[1pt]
\end{tabular}
 \end{table*}

\subsection{Results}\label{sec2:2}

\begin{figure*}[htbp]
\centering
\subfigure
{
\begin{minipage}[t]{0.45\linewidth}
\centering
\includegraphics[width=1\textwidth]{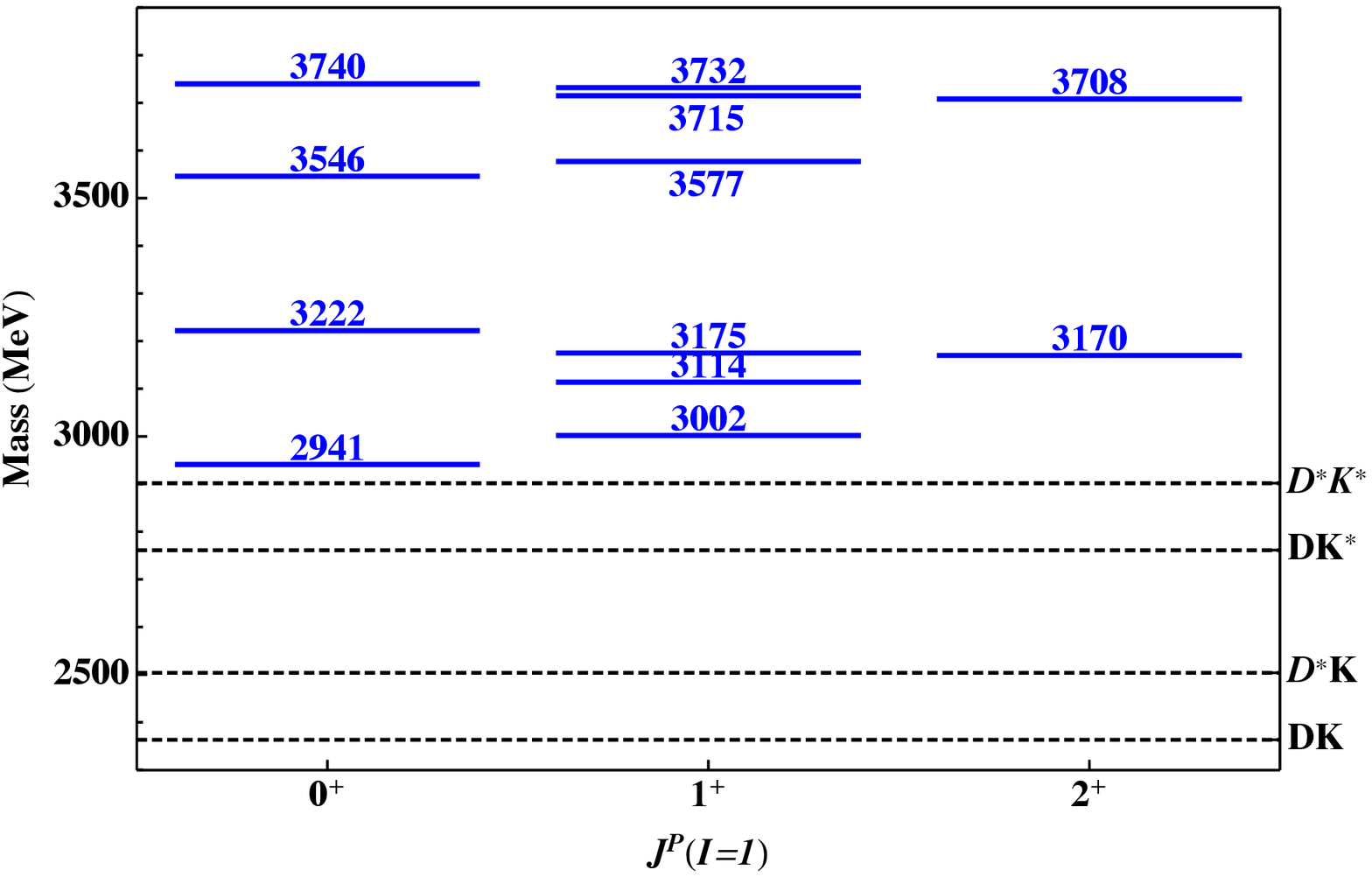}

\end{minipage}%
}%
\subfigure
{
\begin{minipage}[t]{0.45\linewidth}
\centering
\includegraphics[width=1\textwidth]{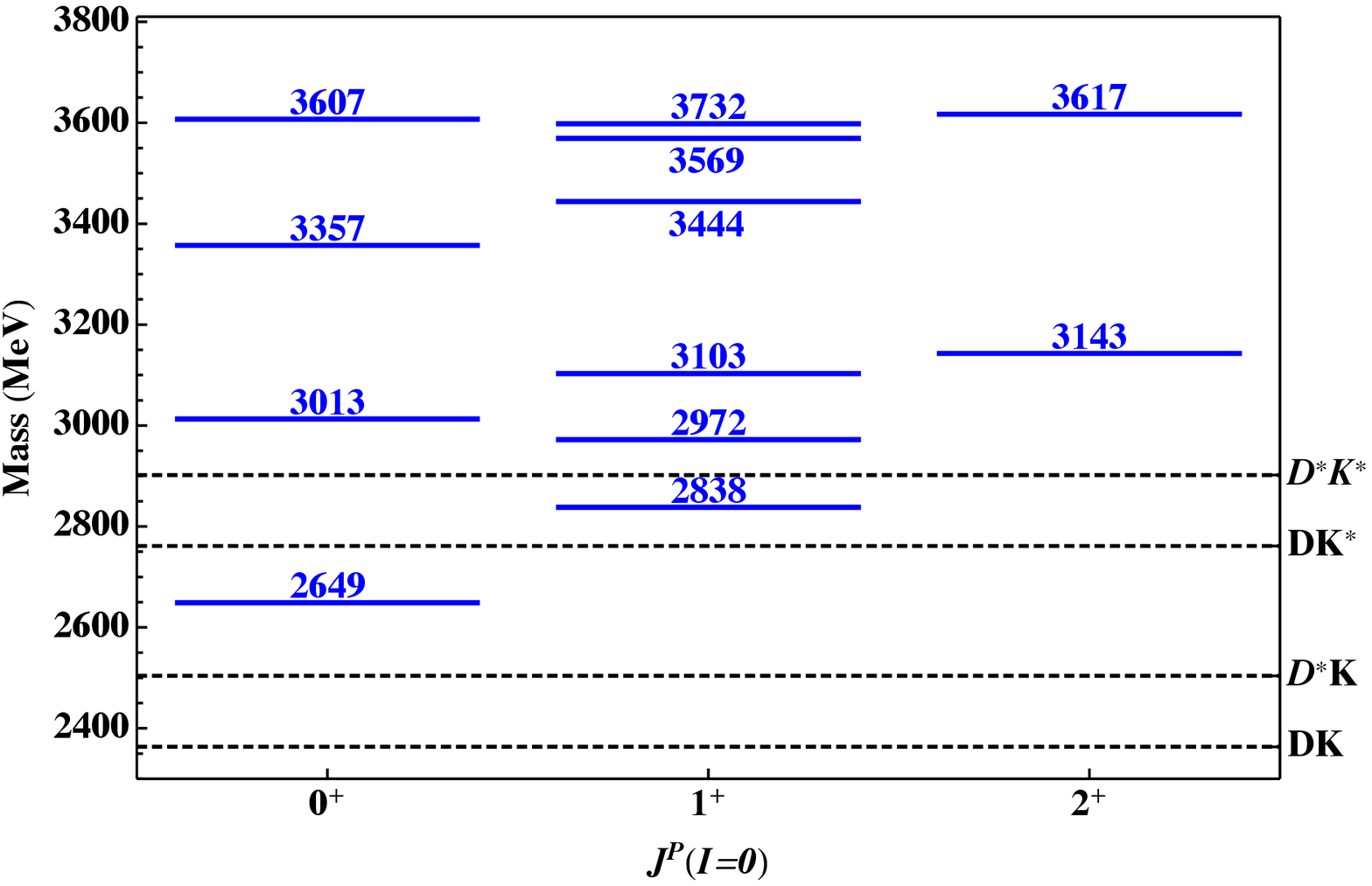}

\end{minipage}%
}%
\centering \caption{The mass spectra of the S-wave tetraquark state
$\bar c\bar s qq$. The dashing lines represent the mass thresholds
of the $\bar D^{(*)}K^{(*)}$ states.} \label{fig:massspectrum}
\end{figure*}

By solving the Schr\"odinger equation with the variational method,
we obtain the mass spectrum and present the results in
Fig.~\ref{fig:massspectrum}. The explicit values of the mass
spectrum and the oscillating parameters are summarized in
Table~\ref{tab:massspectrum}. In Fig.~\ref{fig:massspectrum}, we
find that the tetraquarks with the isospin $I=1$ locate higher than
those with $I=0$. Especially, the ground $I(J^P)=1(0^+)$ state is
about $300$ MeV heavier than the $I(J^P)=0(0^+)$ one. This mass
difference is about $200$ MeV in Ref.~\cite{Lu:2020qmp}. As
illustrated in Table~\ref{csfwavefunction}, the $I(J^P)=0(0^+)$ and
$I(J^P)=1(0^+)$ states contain the same color structures but
different spin-flavor configurations. Thus, with the color
interactions only, i.e., the Coulomb and linear confinement
potentials, the mass spectra of the two states should be the same.
The mass difference comes from the contribution of the hyperfine
potential in view of different spin wave functions. The quark model
which we adopted in this work successfully described the large mass
splitting between the $\pi$ and $\rho$ mesons, which indicates that
the hyperfine interaction should be very important for the light
quarks. In the $\bar c\bar s qq$ system, the significant hyperfine
potential for the diquark $qq$ leads to the large mass splitting
between the $I(J^P)=1(0^+)$ and $I(J^P)=0(0^+)$ states.

\begin{table*}
 \renewcommand\arraystretch{1.8}
 \caption{The mass spectra (in units of MeV) of the tetraquark states $\bar c\bar s qq$ with different $J^P$. The $\beta_{(a/b)}$, $\gamma_{(a/b)}$ and $\lambda_{(a/b)}$ are the oscillating parameters for different states as listed in Table~\ref{csfwavefunction}. In this work, we use $N=2^3$ bases to expand the wave functions of the tetraquark state $\bar c\bar s qq$. Here, we list the results for the S-wave ground and the first radially excited states. }
 \label{tab:massspectrum}
 \centering
 \setlength{\tabcolsep}{2.5mm}
\begin{tabular}{c|c|c|c|c}
\toprule[1pt] \multicolumn{1}{c|}{$J^{P}$} & $I=1$ & Mass & $I=0$ &
Mass\tabularnewline \midrule[1pt] \multirow{4}{*}{$0^{+}$} &
\multirow{2}{*}{$\beta_{a}=0.35$, $\beta_{b}=0.25$, $\beta=0.40$} &
$2941$ & \multirow{2}{*}{$\beta_{a}=0.38$, $\beta_{b}=0.29$,
$\beta=0.43$} & $2649$\tabularnewline
 & & $3222$ & & $3013$\tabularnewline
 & \multirow{2}{*}{$\gamma_{a}=0.33$, $\gamma_{b}=0.25$, $\gamma=0.48$} & $3546$ & \multirow{2}{*}{$\gamma_{a}=0.36$, $\gamma_{b}=0.27$, $\gamma=0.52$} & $3357$\tabularnewline
 & & $3740$ & & $3607$\tabularnewline
\hline \multirow{6}{*}{$1^{+}$} & \multirow{2}{*}{$\beta_{a}=0.36$,
$\beta_{b}=0.24$,$\beta=0.38$} & $3002$ &
\multirow{2}{*}{$\beta_{a}=0.35$, $\beta_{b}=0.28$, $\beta=0.38$} &
$2838$\tabularnewline
 & & $3114$ & & $2972$\tabularnewline
 & \multirow{2}{*}{$\alpha_{a}=0.35$, $\alpha_{b}=0.24$, $\alpha=0.37$} & $3175$ & \multirow{2}{*}{$\alpha_{a}=0.35$, $\alpha_{b}=0.25$, $\alpha=0.49$} & $3103$\tabularnewline
 & & $3577$ & & $3444$\tabularnewline
 & \multirow{2}{*}{$\lambda_{a}=0.32$, $\lambda_{b}=0.24$, $\lambda=0.47$} & $3715$ & \multirow{2}{*}{$\lambda_{a}=0.35$, $\lambda_{b}=0.25$, $\lambda=0.49$} & $3569$\tabularnewline
 & & $3732$ & & $3598$\tabularnewline
\hline \multirow{2}{*}{$2^{+}$} & \multirow{2}{*}{$\beta_{a}=0.32$,
$\beta_{b}=0.23$, $\beta=0.35$} & $3170$ &
\multirow{2}{*}{$\beta_{a}=0.30$, $\beta_{b}=0.22$,
$\text{\ensuremath{\beta}}=0.45$} & $3143$\tabularnewline
 & & $3708$ & & $3617$\tabularnewline
\bottomrule[1pt]
\end{tabular}
\end{table*}

\begin{figure*}[htbp]
\centering \subfigure{
\begin{minipage}[t]{0.38\linewidth}
\centering
\includegraphics[width=1\textwidth]{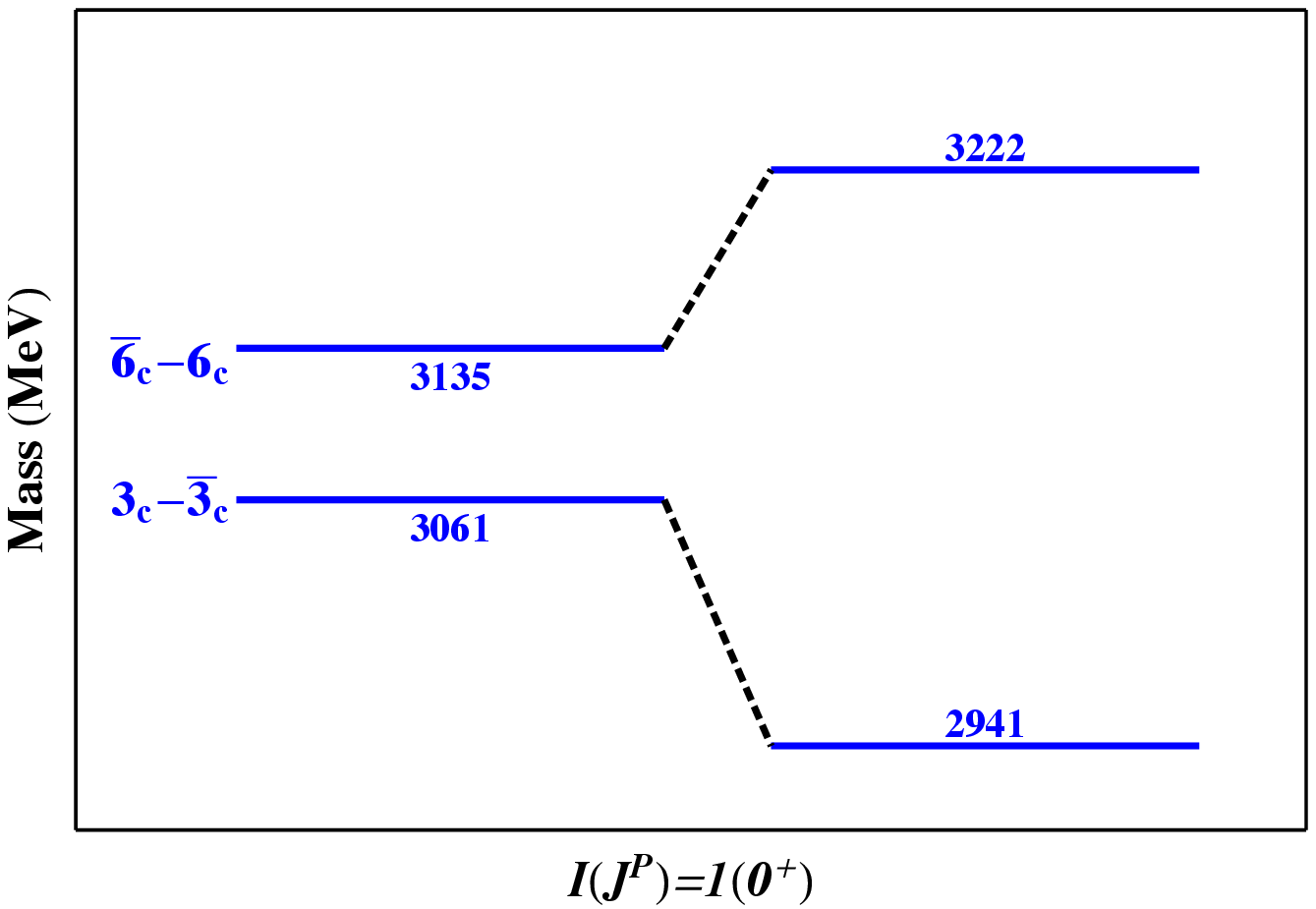}

\end{minipage}%
}%
\subfigure{
\begin{minipage}[t]{0.38\linewidth}
\centering
\includegraphics[width=1\textwidth]{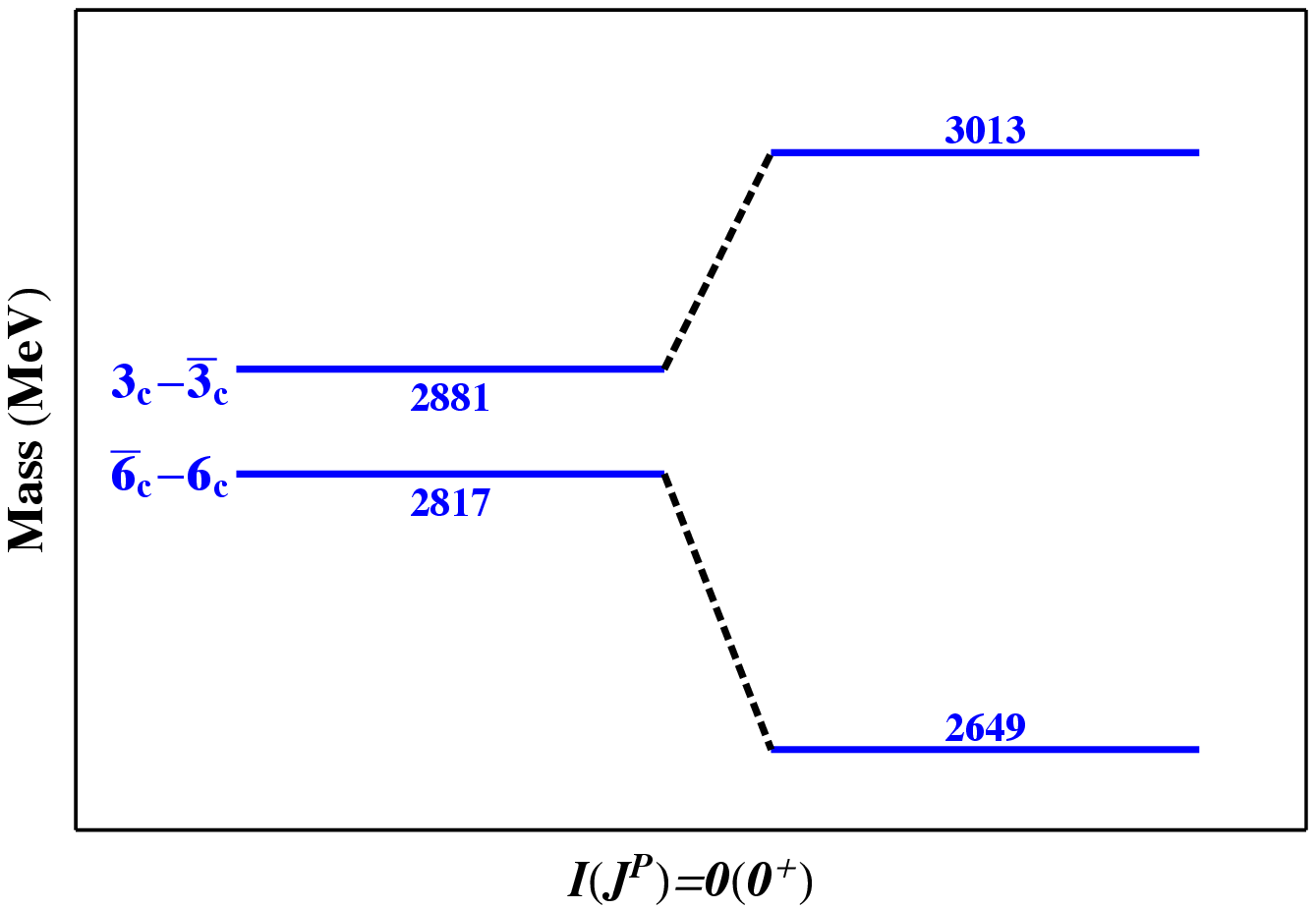}

\end{minipage}%
}%
\centering \caption{The mass spectrum of the $J^P=0^+$ $\bar c\bar s
qq$ state without and with considering the mixing effects between
the$[(\bar c\bar s)_{3_c} (qq)_{\bar 3_c}]_{1_c}$ and $[(\bar c\bar
s)_{6_c} (qq)_{ 6_c}]_{1_c}$ configurations. } \label{fig:mix}
\end{figure*}

In Table~\ref{csfwavefunction}, the $I(J^P)=0(0^+)$ and
$I(J^P)=1(0^+)$ tetraquark states contain two color configurations
$[(\bar c\bar s)_{3_c} (qq)_{\bar 3_c}]_{1_c}$ and $[(\bar c\bar
s)_{6_c} (qq)_{ 6_c}]_{1_c}$, respectively. We first discuss the
mass spectra of the tetraquark states without considering the
coupling between the $ 3_c-\bar 3_c$ and $\bar 6_c- 6_c$ states and
display the results in Fig.~\ref{fig:mix}. For the $I=1$ state, the
$\bar 6_c- 6_c$ state is located higher than the ${ 3_c}-\bar 3_c$
one, while the situation is reversed for the $I=0$ state.

With the mixing effect between the $ 3_c-\bar 3_c$ and $\bar 6_c-
6_c$ states, we obtain the mass spectrum and present them in the
right panel of Fig.~\ref{fig:mix}. The mixing effect comes from the
quark-antiquark interactions between the diquark and the antiquark.
With the quark model in Eq.~\ref{qm}, the coupling constants in the
Coulomb and linear confinement interactions are flavor-independent.
Thus the two interactions cancel with each other exactly and do not
contribute to the color configuration mixing ~\cite{Wang:2019rdo}.
Only the hyperfine interaction contributes. This interaction is
inversely proportional to the masses of the interacting quarks. In
Fig.~\ref{fig:mix}, one finds that the mixing effects shift the
masses at the order of $100$ MeV, which is much larger than those in
the fully heavy tetraquark states~\cite{Wang:2019rdo}.

\begin{table*}
 \renewcommand\arraystretch{1.8}
 \caption{The proportions of different color-spin configurations in the $J^P=0^+$ $\bar c\bar s qq$ system. }
 \label{tab:J0proportion}
 \centering
\begin{tabular}{c|ccccccc}
\toprule[1pt] $J^{P}=0^{+}$ & Mass & $\bar{3}_{c}\otimes3_{c}$ &
$6_{c}\otimes{6}_{c}$ & $\ensuremath{|[(\bar c {q})_{1_{c}}^{0}(\bar
s{q})_{1_{c}}^{0}]_{1_{c}}^{0}\rangle}$ & $\ensuremath{|[(\bar
c{q})_{1_{c}}^{1}(\bar s{q})_{1_{c}}^{1}]_{1_{c}}^{0}\rangle}$ &
$\ensuremath{|[(\bar c {q})_{8_{c}}^{0}(\bar
s{q})_{8_{c}}^{0}]_{1_{c}}^{0}\rangle}$ & $\ensuremath{|(\bar c
{q})_{8_{c}}^{1}(\bar s{q})_{8_{c}}^{1}\rangle}$\tabularnewline
\midrule[1pt] \multirow{2}{*}{$I=1$} & $2941$ & $62.5\%$ & $37.5\%$
& $41.0\%$ & $4.8\%$ & $15.3\%$ & $38.9\%$\tabularnewline
 & $3222$ & $36.8\%$ & $63.2\%$ & $3.1\%$ & $51.3\%$ & $40.3\%$ & $5.3\%$\tabularnewline
\hline \multirow{2}{*}{$I=0$} & $2649$ & $40.2\%$ & $59.8\%$ &
$52.5\%$ & $0.8\%$ & $2.4\%$ & $44.3\%$\tabularnewline
 & $3013$ & $59.9\%$ & $40.1\%$ & $8.5\%$ & $38.2\%$ & $36.5\%$ & $16.8\%$\tabularnewline
\bottomrule[1pt]
\end{tabular}
\end{table*}

\begin{table*}
 \caption{The proportions of different color configurations in the $\bar c\bar s qq$ states with $J^P=1^+$. The $\Psi_1$, $\Psi_2$, and $\Psi_3$ represent the color-flavor-spin configurations in Table~\ref{csfwavefunction}. }
 \centering
 \label{tab:J1proportion}
\begin{tabular}{c|cccccccccc}
\toprule[1pt] $J^{P}=1^{+}$ & Mass & $\Psi_1$ & $\Psi_2$ & $\Psi_3$
& $[(\bar c {q})_{1_{c}}^{0}(\bar s {q})_{1_{c}}^{1}]_{1_{c}}^{1}$ &
$[(\bar c {q})_{1_{c}}^{1}(\bar s{q})_{1_{c}}^{0}]_{1_{c}}^{1}$ &
$[(\bar c {q})_{1_{c}}^{1}(\bar s{q})_{1_{c}}^{1}]_{1_{c}}^{1}$ &
$[(\bar c{q})_{8_{c}}^{0}(\bar s{q})_{8_{c}}^{1}]_{1_{c}}^{1}$ &
$[(\bar c{q})_{8_{c}}^{1}(\bar s{q})_{8_{c}}^{0}]_{1_{c}}^{1}$ &
$[(\bar c{q})_{8_{c}}^{1}(\bar
s{q})_{8_{c}}^{1}]_{1_{c}}^{1}$\tabularnewline \midrule[1pt]
\multirow{3}{*}{$I=1$} & $3002$ & $16.2\%$ & $45.9\%$ & $37.9\%$ &
$7.9\%$ & $36.4\%$ & $1.7\%$ & $1.9\%$ & $11.9\%$ &
$40.2\%$\tabularnewline
 & $3114$ & $79.9\%$ & $18.9\%$ & $1.2\%$ & $28.0\%$ & $4.1\%$ & $1.6\%$ & $44.4\%$ & $13.5\%$ & $8.4\%$\tabularnewline
 & $3175$ & $3.7\%$ & $35.1\%$ & $61.2\%$ & $6.3\%$ & $2.7\%$ & $44.7\%$ & $11.6\%$ & $31.2\%$ & $3.5\%$\tabularnewline
\hline \multirow{3}{*}{$I=0$} & $2838$ & $54.1\%$ & $29.0\%$ &
$16.9\%$ & $0.8\%$ & $46.6\%$ & $1.2\%$ & $15.7\%$ & $1.3\%$ &
$34.3\%$\tabularnewline
 & $2972$ & $30.7\%$ & $68.4\%$ & $0.9\%$ & $43.9\%$ & $9.3\%$ & $3.2\%$ & $3.6\%$ & $27.4\%$ & $12.6\%$\tabularnewline
 & $3103$ & $16.0\%$ & $2.7\%$ & $81.3\%$ & $14.2\%$ & $3.8\%$ & $43.3\%$ & $21.8\%$ & $11.5\%$ & $5.3\%$\tabularnewline
\bottomrule[1pt]
\end{tabular}
\end{table*}

\begin{table*}
 \caption{The proportions of different color configurations in the $cs\bar q \bar q$ states with $J^P=2^+$. }
 \label{tab:J2proportion}
 \centering
 \setlength{\tabcolsep}{2.5mm}
\begin{tabular}{c|ccccccccc|c|cc}
\toprule[1pt]
 I & Mass & $\left[(\bar c \bar s)_{{3}_{c}}^{1}\{ {q}_1 {q}_2\}_{\bar 3_{c}}^{1}\right]_{1_{c}}^{2}$ & $\left[(\bar c\bar s)_{\bar 6_{c}}^{1}[{q}_1{q}_2]_{{6}_{c}}^{1}\right]_{1_{c}}^{2}$ & $\ensuremath{|(\bar c {q})^1_{1_{c}}(\bar s {q})^1_{1_{c}}\rangle}$ & $\ensuremath{|(\bar c {q})^1_{8_{c}}(\bar s{q})^1_{8_{c}}\rangle}$ \tabularnewline
\midrule[1pt] \multirow{2}{*}{$I=1$} & $3170$ & $100\%$ & $0$ &
$33.3\%$ & $66.6\%$ \tabularnewline
  & $3708$ & $100\%$ & $0$ & $33.3\%$ & $66.6\%$ \tabularnewline
\hline \multirow{2}{*}{$I=0$} & $3143$ & $0$ & $100\%$ & $66.6\%$ &
$33.3\%$ \tabularnewline
  & $3617$ & $0$ & $100\%$ & $66.6\%$ & $33.3\%$ \tabularnewline
\bottomrule[1pt]
\end{tabular}
\end{table*}

We also investigate the inner structures of the tetraquark states,
including the proportions of different color configurations as
listed in Tables~\ref{tab:J0proportion}-~\ref{tab:J2proportion} and
the root mean square radii between different interacting quarks in
Table~\ref{tab:radius}. The proportions show that the ground S-wave
states contain important color configuration $[(\bar c q)_{8_c}(\bar
s q)_{8_c}]_{1_c}$. As illustrated in Table~\ref{tab:radius},
almost all of the root mean square radii are smaller than $1$ fm,
which indicates that the four quarks are compactly bound within the
tetraquark states. One should be careful in identifying the states as the resonances	.~Since we have used the finite numbers of the bases to expand the tetraquark state, the eigenvectors may correspond to the bound or the resonance  states, and the continuum in the finite space. To distinguish the resonance from the continuum, we may include their coupling and investigate whether the tetraquark states may remain as resonances or not by the real scaling method in the future \cite{Hiyama:2003cu,Meng:2020knc}.

The mass of the $I(J^P)=1(0^+)$ state is $2941$ MeV, which is $75$
MeV larger than that of the $X_0(2900)$ state. Such a mass deviation
is not very significant and conclusive if one considers the
uncertainty of the quark model. One notices that the model
parameters used in the mass spectrum calculation are obtained from
the conventional hadrons. As a phenomenological model, the
parameters may change for different systems. For instance, in
Refs.~\cite{Godfrey:1985xj} and~\cite{Capstick:1986bm}, the authors
studied the mass spectra of the mesons and the baryons in a unified
relativized quark model. The overall constant term in the
meson-based potential was $-340$ MeV , while the baryon-based one
was selected as $-615$ MeV to obtain better solutions for the
baryons $qqq$. When the meson-based quark model is applied to the
multi-quark systems, the similar modification might also be
essential. Furthermore, the confinement mechanisms in the
multi-quark systems are still poorly understood. For example,
because of the rich color configurations in the multi-quark system,
the three-body interaction arising from the three-gluon vertex may
also contribute, while it vanishes for the conventional hadrons.
Considering the uncertainties of quark model, the $X(2900)$ signal
seems a candidate of the $I(J^P)=1(0^+)$ state. Besides the
$X(2941)$, we also predict the other states such as the
$I(J^P)=0(0^+)$ state with the mass around $2649$ MeV.

\begin{table*}
 \caption{The root mean square radii (in units of fm) between different interacting quarks in the $cs\bar q \bar q$ states with $J^P=0^+$, $J^P=1^+$ and $J^P=2^+$. }
 \label{tab:radius}
 \centering
 \setlength{\tabcolsep}{2.5mm}
\begin{tabular}{c|ccccccccc}
\toprule[1pt] \multicolumn{1}{c|}{$J^{P}=0^{+}$} & Mass & $r_{\bar
c\bar s}$ & $r_{{q}{q}}$ & $r$ & $r_{\bar c{q}}$ & $r_{\bar s{q}}$ &
$r'$ \tabularnewline \midrule[1pt] \multirow{2}{*}{$I=1$} & $2941$ &
$0.64$ & $0.91$ & $0.51$ & $0.70$ & $0.84$ & $0.51$ \tabularnewline
 & $3222$ & $0.76$ & $1.03$ & $0.56$ & $0.79$ & $0.96$ & $0.60$ \tabularnewline
\hline \multirow{2}{*}{$I=0$} & $2649$ & $0.64$ & $0.85$ & $0.48$ &
$0.61$ & $0.74$ & $0.46$ \tabularnewline
 & $3013$ & $0.71$ & $0.94$ & $0.61$ & $0.72$ & $0.86$ & $0.51$ \tabularnewline
 \midrule[1pt]
 \multicolumn{1}{c|}{$J^{P}=1^{+}$} & Mass & $r_{\bar c\bar s}$ & $r_{{q}{q}}$ & $r$ & $r_{\bar c {q}}$ & $r_{\bar s{q}}$ & $r'$\tabularnewline
 \midrule[1pt]
\multirow{3}{*}{$I=1$} & $3002$ & $0.64$ & $0.91$ & $0.48$ & $0.74$
& $0.87$ & $0.52$\tabularnewline
 & $3114$ & $0.64$ & $0.94$ & $0.59$ & $0.79$ & $0.91$ & $0.52$\tabularnewline
 & $3175$ & $0.74$ & $1.03$ & $0.55$ & $0.79$ & $0.96$ & $0.59$\tabularnewline
\hline \multirow{3}{*}{$I=0$} & $2838$ & $0.63$ & $0.82$ & $0.52$ &
$0.49$ & $0.68$ & $0.81$\tabularnewline
 & $2972$ & $0.71$ & $0.89$ & $0.52$ & $0.55$ & $0.70$ & $0.87$\tabularnewline
 & $3103$ & $0.75$ & $1.03$ & $0.58$ & $0.59$ & $0.74$ & $0.92$\tabularnewline
\midrule[1pt] $J^{P}=2^{+}$ & Mass & $r_{\bar c\bar s}$ &
$r_{{q}{q}}$ & $r$ & $r_{\bar c {q}}$ & $r_{\bar s {q}}$ &
$r'$\tabularnewline
 \midrule[1pt]
\multirow{2}{*}{$I=1$} & $3170$ & $0.66$ & $0.97$ & $0.65$ & $0.83$
& $0.96$ & $0.54$\tabularnewline
 & $3708$ & $0.88$ & $0.98$ & $0.74$ & $0.89$ & $1.10$ & $0.68$\tabularnewline
\hline \multirow{2}{*}{$I=0$} & $3863$ & $0.83$ & $1.04$ & $0.51$ &
$0.76$ & $0.97$ & $0.64$\tabularnewline
 & $4109$ & $0.93$ & $1.05$ & $0.51$ & $0.77$ & $1.02$ & $0.71$\tabularnewline
\bottomrule[1pt]
\end{tabular}
\end{table*}

\section{Decay width}\label{sec3}

In this section, we calculate the strong decay widths of the
teraquark states $\bar c\bar sq q$. The dominant decay modes of
tetraquark states $\bar c\bar sq q$ are $X\to\bar D^{(*)} K^{(*)}$
when the phase spaces are allowed. The tetraquarks may also decay
into the $\bar D K^{(*)} \pi$ final state. The three-body decays are
largely suppressed compared with the two-body decays due to the
phase space suppression. Thus, we do not consider them in this work.

\subsection{Quark interchange model}
\begin{figure*}[htbp]
\centering
\includegraphics[width=0.8\textwidth]{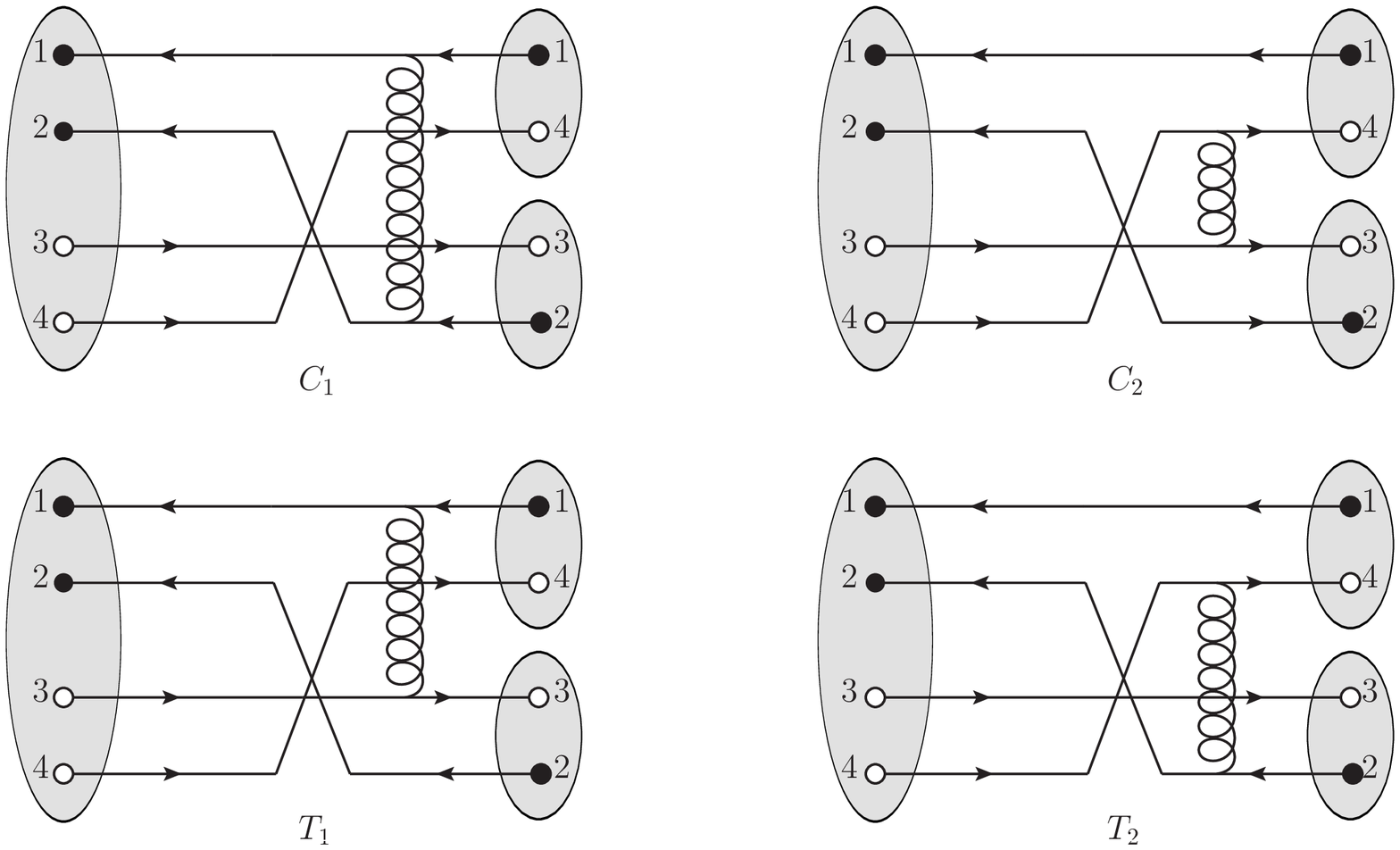}
\caption{The diagrams for the tetraquark decaying into two mesons at
the quark level. The $1-4$ denote the $\bar c$, $\bar s$, $q$, $q$
quarks, respectively. The curve line denotes the quark-quark
interactions.} \label{Fig:decay}
\end{figure*}

We first give a brief introduction to the quark-interchange model
for the decay
\begin{eqnarray}
\bar c \bar s q q\rightarrow B(c\bar q)+C(s\bar q),
\end{eqnarray}
where $B$ and $C$ are two color singlet mesons. The decay process
occurs through interchanging the constituent quarks (antiquarks) as
illustrated in Fig.~\ref{Fig:decay}, followed by hadronization into
two mesons. This method has been applied to calculate the $I=3/2$
$K\pi$ scattering~\cite{Barnes:1992qa}, the short-range NN
interaction~\cite{Barnes:1993nu}, the strong decays of the $Z_c$
states~\cite{Xiao:2019spy}, X(3872)~\cite{Zhou:2019swr}, and the
pentaquark states~\cite{Wang:2019spc}.
 In the decay, the interacting Hamiltonian reads
\begin{eqnarray}
H_{ij}^I(r_{ij})=\sum_{i<j}V_{ij}(r_{ij}),
\end{eqnarray}
where $V_{ij}$ is the potential as listed in Eq. (\ref{qm}).  The
$T$-matrix element is the sum of the four diagrams in
Fig.~\ref{Fig:decay}, which are calculated as the overlaps of the
wave functions with the Hamiltonian between the initial and final
states. The wave functions of the initial tetraquark states and
final mesons are obtained from the same quark model in
Eq.~\eqref{para}. For each diagram, the $T_{fi}$ is written as the
product of factors (as defined in Ref.~\cite{Xiao:2019spy})
\begin{eqnarray}
T_{fi} &=&I_{\text {flavor}}I_{\text{color}} I_{\text{spin-space}}.
\end{eqnarray}
The element $I_{\text {flavor}}$ is the overlap of wave functions in
the flavor space. The color and the spin matrix elements are listed
in Table~\ref{tab:decayfactor} in Appendix. The calculation details
of the spatial matrix element $I_{\text {space}}$ are referred to
Ref.~\cite{Xiao:2019spy}.

The differential decay width is given by
\begin{eqnarray}
\frac{d\Gamma}{d\Omega}=\frac{1}{2J+1}\frac{|{\vec
p}_B|}{32\pi^2M^2}|\mathcal M (\bar c \bar s q q\rightarrow B(\bar
cq)+C(\bar sq) )|^2,
\end{eqnarray}
where ${\vec p}_B$ is the momentum of the final state. $M$ is the
mass of the initial tetraquark state. The decay amplitude $\mathcal
M$ is related to the T-matrix $T_{fi}$ by
\begin{eqnarray}
\mathcal M(\bar c \bar s\bar q\bar q\rightarrow B+C)=-(2\pi)^{3/2}\sqrt{2M}\sqrt{2E_B}\sqrt{2E_C}T_{fi}, \nonumber \\
\end{eqnarray}
where the $E_B$ and $E_C$ are the energies of the two mesons in the
final state.

\subsection{Numerical results}
We present the decay widths of the tetraquark states in
Table~\ref{tab:decay width}. The $J^{P}=0^{+}$ states are of special
interest because they might be the candidates for the recently
observed $X_{0}(2900)$ in the LHCb Collaboration. Two S-wave decay
modes are available to the states, $\bar{D} {K}$ and
$\bar{D}^{*}K^{*}$. For the $X(2941)$ and the $X(2649)$, we predict
the partial decay widths into the $\bar{D}K$ channel as $20.1$ MeV
and $48.1$ MeV, respectively. And the partial width ratio is
\begin{eqnarray}
\frac{\Gamma_{DK}(X(2941))}{{\Gamma_{DK}}(X(2649))}=0.4.
\end{eqnarray}
The $X(2941)$ also decays into the $\bar{D}^{*}K^{*}$ with a smaller
decay width $6.5$ MeV, while the mode is energetically forbidden for
the $X(2649)$. In Sec.~\ref{sec2:2}, the mass of $X(2941)$ state is
larger than the $X_0(2900)$ mass by $75$ MeV. The predicted decay
width $26.6$ MeV is reasonably close to the experimental value of
$57.2\pm12.9$ MeV. If we regard the $X(2941)$ as the $X_0(2900)$
state and use the experimental value as the input mass in the decay,
the $\bar{D}^{*}K^{*} $ decay mode is forbidden. It decays into the
$\bar{D}K$ channel with the decay width $21.7$ MeV. The partial
width ratio is
\begin{eqnarray}
\frac{{\Gamma_{\bar{D}K}}(X(2900))}{{\Gamma_{\bar{D}K}}(X(2649))}=0.5.
\end{eqnarray}
The ratio indicates that if the $X_0(2900)$ is the
$I(J^{P})=1(0^{+})$ state, there might exist the other
$I(J^P)=0(0^+)$ state $X(2649)$ in the $\bar{D}K$ channel. More
experimental study is expected in future to test the tetraquark
explanation for the $X_0(2900)$.

For the $J^P=0^+$ states, the  $X(3222)$ and $X(3013)$ states are
promising to be observed in $\bar D^*K^*$ mode. The $\bar{D}K$
widths of the two states are much smaller than those of the $\bar
D^*K^*$ ones, despite the larger phase spaces. The
${{\Gamma_{DK}}}(X(3222))$ is predicted to be $0.01$ MeV, as it
weekly couples with the $\bar{D}K$ channel. The $J^P=2^+$ states
also decay into the $\bar D^*K^*$ mode. However, the partial decay
widths are rather small and these tetraquark states may be difficult
to be observed in this mode.

The $J^{P}=1^{+}$ states may decay into the $\bar D^* K$, $\bar D
K^*$ and $\bar D^* K^*$ channels. Most of the predicted widths are
much smaller than those of the $J^{P}=0^{+}$ states except the
$X(2838)$ and $X(2972)$. They both have large partial decay widths
into the $\bar D^* K$ and $\bar D K^*$, respectively.

\begin{table*}
 \caption{The decay widths (in units of MeV) for the tetraquark states $\bar c\bar s qq$ decaying into the $\bar D^{(*)}K^{(*)}$. The script ``$-$" presents that the corresponding decay mode is forbidden by the phase space. }\label{tab:decay width}
\begin{tabular}{c|cccc|c|ccccc|c|cc}
\toprule[1pt] \multicolumn{5}{c|}{$J^{P}=0^{+}$} &
\multicolumn{6}{c|}{$J^{P}=1^{+}$} &
\multicolumn{3}{c}{$J^{P}=2^{+}$}\tabularnewline \hline
 & Mass & $\Gamma_{X\rightarrow \bar DK}$ & $\Gamma_{X\rightarrow \bar D^{*}K^{*}}$ & $\Gamma$ & & Mass & $\Gamma_{X\rightarrow \bar D^{*}K}$ & $\Gamma_{X\rightarrow \bar DK^{*}}$ & $\Gamma_{X\rightarrow \bar D^{*}K^{*}}$ & ${{\Gamma}}$ & \multicolumn{1}{c|}{} & \multicolumn{1}{c}{Mass} & $\Gamma_{X\rightarrow \bar D^{*}K^{*}}$\tabularnewline
\midrule[1pt] 
\multirow{2}{*}{$I=1$} & $2941$ & $20.1$ & $6.5$ & $26.6$ &
\multirow{3}{*}{$I=1$} & $3002$ & $5.1$ & $1.6$ & $0.2$ & $6.9$ &
\multicolumn{1}{c|}{$I=1$} & \multicolumn{1}{c}{$3170$} & $0.3$
\tabularnewline
 \cline{12-14}
 & $3222$ & $0.01$ & $26.4$ & $26.4$ & & $3114$ & $3.1$ & $2.3$ & $1.7$ & $7.1$ & \multicolumn{1}{c|}{$I=0$} & \multicolumn{1}{c}{$3143$} & $0.1$ \tabularnewline
\cline{1-5} 
\multirow{2}{*}{$I=0$} & $2649$ & $48.1$ & - & $48.1$ & & $3175$ &
$1.3$ & $7.3$ & $3.5$ & {$12.1$} & & &
\multicolumn{1}{c}{}\tabularnewline
\cline{6-11} 
 & $3013$ & $15.9$ & $88.5$ & $104.4$ & \multirow{3}{*}{$I=0$} & $2838$ & $28.1$ & $3.0$ & $-$ &{$31.1$} & & & \multicolumn{1}{c}{}\tabularnewline
 & \multicolumn{1}{c}{} & \multicolumn{1}{c}{} & \multicolumn{1}{c}{} & & & $2972$ & $1.8$ & $20.0$ & $13.0$ &{$34.8$} & & & \multicolumn{1}{c}{}\tabularnewline
 & \multicolumn{1}{c}{} & \multicolumn{1}{c}{} & \multicolumn{1}{c}{} & & & $3103$ & $3.5$ & $4.2$ & $11.0$ & {$18.7$} & & & \multicolumn{1}{c}{}\tabularnewline
\bottomrule[1pt]
\end{tabular}
\end{table*}

\section{Summary}\label{sec4}

In this work, we evaluate the mass spectrum and the decay widths of
the system $cs\bar q \bar q$ with the quark model. We include the
most important ingredients of the quark model, the Coulomb, the
linear confinement, and the hyperfine interactions. With the
results, we examine whether the $X_0(2900)$ can be interpreted as a tetraquark state.

We first solve the Schr\"odinger equation with the variational method.
We obtain the mass spectrum of the S-wave $\bar c \bar s qq$ states
with $J^P=0^+$, $1^+$ and $2^+$. The $J^P=0^+$ and $1^+$ states are
the mixtures of different color-spin-flavor configurations. In the
quark model, the color interaction does not mix different color-spin
configurations. Only the hyperfine interactions contribute to their
mixing effects. Since the $\bar c\bar s qq$ state contains the light
quark, the mixing effect is larger than that in the fully heavy
tetraquark state. With the Fierz transformation, we obtain the proportions
of the $(\bar c q)_{1_c}(\bar s q)_{1_c}$ and $(\bar c q)_{8_c}(\bar s q)_{8_c}$ components. The ground state contains large color component $(\bar c q)_{8_c}(\bar s q)_{8_c}$. For the $J^P=0^+$ $\bar c \bar s qq$ system, the two isospin partner states have the same color configurations but different color-spin configurations. 
The mass splittings come from the hyperfine interactions.

With the obtained wave functions from the mass spectrum calculation,
we study the strong decays for the tetraquarks $\bar c\bar sqq$ into
the $\bar D^{(*)}K^{(*)}$ channels using the quark interchange
model. The partial decay widths might be helpful in the search for
tetraquark states $\bar c\bar s qq$, as they suggest the golden
channel to reconstruct the tetraquark states.

Combined the mass spectrum and decay width, the $X(2941)$ with the
quantum number $I(J^P)=1(0^+)$ seems to be a good candidate for the
$X_0(2900)$. Its mass and decay width are $M=2941$ MeV and
$\Gamma=26.6$ MeV, respectively. The partial decay widths into the
$\bar D K$ and $\bar D^* K^*$ modes are $20.1$ and $6.5$ MeV. If the
$X(2941)$ is treated as $X_0(2900)$, the $\bar D^* K^*$ mode is
kinematically forbidden and the $\bar D K$ width is $21.7$ MeV,
which is close to the experimental data. Remarkably, we also find
another state $I(J^P)=0(0^+)$ with the $M=2649$ MeV and
$\Gamma_{X\rightarrow DK}=48.1$ MeV at the same time. The decay
width ratio $\Gamma(X(2941)\rightarrow DK)/\Gamma(X(2649)\rightarrow
DK)=0.4$ indicates that if the $X(2941)$ is regarded as the
$X(2900)$, there might be another $X(2649)$ state in the $\bar D K$
channel. In addition, one may also expect other states to be
observed in the $\bar D ^*K$, $\bar D^* K$, and $\bar D^* K^*$
channels. The possible observation channels in experiments are
related to their partial decay widths.

So far, we have evaluated the mass spectrum and the decay width with
a meson-based quark model. The potentials in the multi-quark system
might be different from the meson-based potentials. For instance,
the parameters might change and the different confinement mechanisms
might show up, which may bring uncertainties to our results and will
be discussed lately. We expect more experimental measurements in
future to help test our results and improve the understanding of the
interaction in the multi-quark system, especially the confinement
mechanisms and dynamics in the strong decay.

\section*{ACKNOWLEDGMENTS}
G. J. Wang is very grateful to X. Z. Weng for very helpful
discussions. This project is supported by the National Natural
Science Foundation of China under Grant 11975033. G.J.Wang is also
supported by China Postdoctoral Science Foundation Grant No.
2019M660279.

\section*{Appendix}
The hyperfine interaction plays an important role in the mass
spectrum calculation, because it contributes to the mixing effects
of different color-spin configurations, as well as inducing the
large mass splitting between the $I(J^P)=1(1^+)$ and
$I(J^P)=0({0^+})$ ground states. We list the color-spin factor
$\langle \frac{\mathbf \lambda_i}{2} \frac{\mathbf
\lambda_j}{2}\mathbf s_i\cdot \mathbf s_j \rangle$ of the hyperfine
interactions in the mass spectrum calculation in Table~\ref
{tab:ss}.
\begin{table*}
 \caption{The factor $\langle \frac{\mathbf \lambda_i}{2} \frac{\mathbf \lambda_j}{2}\mathbf s_i\cdot \mathbf s_j \rangle$ for the $\bar c\bar s q_1q_2$ state in the mass spectrum calculation.  The subscripts of the ``$\langle...\rangle$" correspond to the indice of the states $\Psi$ in Table~\ref{csfwavefunction}. }\label{tab:ss}
\begin{tabular}{c|cccccc|cccccc}
\toprule[1pt] $J^{P}$ & \multicolumn{6}{c|}{$I=1$} &
\multicolumn{6}{c}{$I=0$}\tabularnewline \midrule[1pt]
\multirow{6}{*}{$0^{+}$} & $\langle \bar cq_1 \rangle_{11}$ &
$\langle \bar s q_2 \rangle_{11}$ & $\langle \bar c q_2
\rangle_{11}$ & $\langle \bar s q_1\rangle_{11}$ & $\langle \bar c
\bar s\rangle_{11}$ & $\langle q_1q_2 \rangle_{11}$ & $\langle \bar
cq_1 \rangle_{11}$ & $\langle \bar s q_2 \rangle_{11}$ & $\langle
\bar c q_2 \rangle_{11}$ & $\langle \bar s q_1\rangle_{11}$ &
$\langle \bar c \bar s\rangle_{11}$ & $\langle q_1q_2
\rangle_{11}$\tabularnewline
 & $\frac{1}{6}$ & $\frac{1}{6}$ & $\frac{1}{6}$ & $\frac{1}{6}$ & $-\frac{1}{6}$ & $-\frac{1}{6}$ & $0$ & $0$ & $0$ & $0$ & $\frac{1}{2}$ & $\frac{1}{2}$\tabularnewline
 & $\langle \bar cq_1 \rangle_{22}$ & $\langle \bar s q_2 \rangle_{22}$ & $\langle \bar c q_2 \rangle_{22}$ & $\langle \bar s q_1\rangle_{22}$ & $\langle \bar c \bar s\rangle_{22}$ & $\langle q_1q_2 \rangle_{22}$ & $\langle \bar cq_1 \rangle_{22}$ & $\langle \bar s q_2 \rangle_{22}$ & $\langle \bar c q_2 \rangle_{22}$ & $\langle \bar s q_1\rangle_{22}$ & $\langle \bar c \bar s\rangle_{22}$ & $\langle q_1q_2 \rangle_{22}$\tabularnewline
 & $0$ & $0$ & $0$ & $0$ & $-\frac{1}{4}$ & $-\frac{1}{4}$ & $\frac{5}{12}$ & $\frac{5}{12}$ & $\frac{5}{12}$ & $\frac{5}{12}$ & $\frac{1}{12}$ & $\frac{1}{12}$\tabularnewline
 & $\langle \bar cq_1 \rangle_{12}$ & $\langle \bar s q_2 \rangle_{12}$ & $\langle \bar c q_2 \rangle_{12}$ & $\langle \bar s q_1\rangle_{12}$ & $\langle \bar c \bar s\rangle_{12}$ & $\langle q_1q_2 \rangle_{12}$   & $\langle \bar cq_1 \rangle_{12}$ & $\langle \bar s q_2 \rangle_{12}$ & $\langle \bar c q_2 \rangle_{12}$ & $\langle \bar s q_1\rangle_{12}$ & $\langle \bar c \bar s\rangle_{12}$ & $\langle q_1q_2 \rangle_{12}$\tabularnewline
 & $\frac{\sqrt{3}}{4\sqrt{2}}$ & $\frac{\sqrt{3}}{4\sqrt{2}}$ & $\frac{\sqrt{3}}{4\sqrt{2}}$ & $\frac{\sqrt{3}}{4\sqrt{2}}$ & $0$ & $0$ & $\frac{\sqrt{3}}{4\sqrt{2}}$ & $\frac{\sqrt{3}}{4\sqrt{2}}$ & $\frac{\sqrt{3}}{4\sqrt{2}}$ & $\frac{\sqrt{3}}{4\sqrt{2}}$ & $0$ & $0$\tabularnewline
\midrule[1pt] \multirow{12}{*}{$1^{+}$} & $\langle \bar cq_1
\rangle_{11}$ & $\langle \bar s q_2 \rangle_{11}$ & $\langle \bar c
q_2 \rangle_{11}$ & $\langle \bar s q_1\rangle_{11}$ & $\langle \bar
c \bar s\rangle_{11}$ & $\langle q_1q_2 \rangle_{11}$ & $\langle
\bar cq_1 \rangle_{11}$ & $\langle \bar s q_2 \rangle_{11}$ &
$\langle \bar c q_2 \rangle_{11}$ & $\langle \bar s q_1\rangle_{11}$
& $\langle \bar c \bar s\rangle_{11}$ & $\langle q_1q_2
\rangle_{11}$\tabularnewline
 & $\frac{1}{12}$ & $\frac{1}{12}$ & $\frac{1}{12}$ & $\frac{1}{12}$ & $-\frac{1}{6}$ & $-\frac{1}{6}$ & $0$ & $0$ & $0$ & $0$ & $-\frac{1}{6}$ & $\frac{1}{2}$\tabularnewline
 & $\langle \bar cq_1 \rangle_{22}$ & $\langle \bar s q_2 \rangle_{22}$ & $\langle \bar c q_2 \rangle_{22}$ & $\langle \bar s q_1\rangle_{22}$ & $\langle \bar c \bar s\rangle_{22}$ & $\langle q_1q_2 \rangle_{22}$ & $\langle \bar cq_1 \rangle_{22}$ & $\langle \bar s q_2 \rangle_{22}$ & $\langle \bar c q_2 \rangle_{22}$ & $\langle \bar s q_1\rangle_{22}$ & $\langle \bar c \bar s\rangle_{22}$ & $\langle q_1q_2 \rangle_{22}$\tabularnewline
 & $0$ & $0$ & $0$ & $0$ & $\frac{1}{2}$ & $-\frac{1}{6}$ & $\frac{5}{24}$ & $\frac{5}{24}$ & $\frac{5}{24}$ & $\frac{5}{24}$ & $\frac{1}{12}$ & $\frac{1}{12}$\tabularnewline
 & $\langle \bar cq_1 \rangle_{33}$ & $\langle \bar s q_2 \rangle_{33}$ & $\langle \bar c q_2 \rangle_{33}$ & $\langle \bar s q_1\rangle_{33}$ & $\langle \bar c \bar s\rangle_{33}$ & $\langle q_1q_2 \rangle_{33}$ & $\langle \bar cq_1 \rangle_{33}$ & $\langle \bar s q_2 \rangle_{33}$ & $\langle \bar c q_2 \rangle_{33}$ & $\langle \bar s q_1\rangle_{33}$ & $\langle \bar c \bar s\rangle_{33}$ & $\langle q_1q_2 \rangle_{33}$\tabularnewline
 & $0$ & $0$ & $0$ & $0$ & $\frac{1}{12}$ & $-\frac{1}{4}$ & $0$ & $0$ & $0$ & $0$ & $-\frac{1}{4}$ & $\frac{1}{12}$\tabularnewline
  & $\langle \bar cq_1 \rangle_{12}$ & $\langle \bar s q_2 \rangle_{12}$ & $\langle \bar c q_2 \rangle_{12}$ & $\langle \bar s q_1\rangle_{12}$ & $\langle \bar c \bar s\rangle_{12}$ & $\langle q_1q_2 \rangle_{12}$   & $\langle \bar cq_1 \rangle_{12}$ & $\langle \bar s q_2 \rangle_{12}$ & $\langle \bar c q_2 \rangle_{12}$ & $\langle \bar s q_1\rangle_{12}$ & $\langle \bar c \bar s\rangle_{12}$ & $\langle q_1q_2 \rangle_{12}$\tabularnewline
 & $\frac{1}{6\sqrt{2}}$ & $-\frac{1}{6\sqrt{2}}$ & $\frac{1}{6\sqrt{2}}$ & $-\frac{1}{6\sqrt{2}}$ & $0$ & $0$ & $-\frac{1}{4}$ & $\frac{1}{4}$ & $-\frac{1}{4}$ & $\frac{1}{4}$ & $0$ & $0$\tabularnewline
 & $\langle \bar cq_1 \rangle_{13}$ & $\langle \bar s q_2 \rangle_{13}$ & $\langle \bar c q_2 \rangle_{13}$ & $\langle \bar s q_1\rangle_{13}$ & $\langle \bar c \bar s\rangle_{13}$ & $\langle q_1q_2 \rangle_{13}$   & $\langle \bar cq_1 \rangle_{13}$ & $\langle \bar s q_2 \rangle_{13}$ & $\langle \bar c q_2 \rangle_{13}$ & $\langle \bar s q_1\rangle_{13}$ & $\langle \bar c \bar s\rangle_{13}$ & $\langle q_1q_2 \rangle_{13}$\tabularnewline
 & $-\frac{1}{4}$ & $\frac{1}{4}$ & $-\frac{1}{4}$ & $\frac{1}{4}$ & $0$ & $0$ & $-\frac{1}{4\sqrt{2}}$ & $-\frac{1}{4\sqrt{2}}$ & $-\frac{1}{4\sqrt{2}}$ & $-\frac{1}{4\sqrt{2}}$ & $0$ & $0$\tabularnewline
& $\langle \bar cq_1 \rangle_{23}$ & $\langle \bar s q_2
\rangle_{23}$ & $\langle \bar c q_2 \rangle_{23}$ & $\langle \bar s
q_1\rangle_{23}$ & $\langle \bar c \bar s\rangle_{23}$ & $\langle
q_1q_2 \rangle_{23}$   & $\langle \bar cq_1 \rangle_{23}$ & $\langle
\bar s q_2 \rangle_{23}$ & $\langle \bar c q_2 \rangle_{23}$ &
$\langle \bar s q_1\rangle_{23}$ & $\langle \bar c \bar
s\rangle_{23}$ & $\langle q_1q_2 \rangle_{23}$\tabularnewline

 & $-\frac{1}{4\sqrt{2}}$ & $-\frac{1}{4\sqrt{2}}$ & $-\frac{1}{4\sqrt{2}}$ & $-\frac{1}{4\sqrt{2}}$ & $0$ & $0$ & $\frac{5}{12\sqrt{2}}$ & $-\frac{5}{12\sqrt{2}}$ & $\frac{5}{12\sqrt{2}}$ & $-\frac{5}{12\sqrt{2}}$ & $0$ & $0$\tabularnewline
\midrule[1pt] \multirow{2}{*}{$2^{+}$} & $\langle \bar cq_1
\rangle_{11}$ & $\langle \bar s q_2 \rangle_{11}$ & $\langle \bar c
q_2 \rangle_{11}$ & $\langle \bar s q_1\rangle_{11}$ & $\langle \bar
c \bar s\rangle_{11}$ & $\langle q_1q_2 \rangle_{11}$ & $\langle
\bar cq_1 \rangle_{11}$ & $\langle \bar s q_2 \rangle_{11}$ &
$\langle \bar c q_2 \rangle_{11}$ & $\langle \bar s q_1\rangle_{11}$
& $\langle \bar c \bar s\rangle_{11}$ & $\langle q_1q_2
\rangle_{11}$\tabularnewline
 & $-\frac{1}{12}$ & $-\frac{1}{12}$ & $-\frac{1}{12}$ & $-\frac{1}{12}$ & $-\frac{1}{6}$ & $-\frac{1}{6}$ & $-\frac{5}{24}$ & $-\frac{5}{24}$ & $-\frac{5}{24}$ & $-\frac{5}{24}$ & $\frac{1}{12}$ & $\frac{1}{12}$\tabularnewline
\bottomrule[1pt]
\end{tabular}
\end{table*}

The values of the color matrix element $I_{\text {color}}$ and the
spin factor $I_{\text{spin}}$ in the decay amplitudes are tabulated
in Table~\ref{tab:decayfactor}.

\begin{table*}
 \caption{The color matrix element $I_{\text {color}}=\langle\frac{\mathbf{\lambda}_{i}}{2}\cdot\frac{\mathbf{\lambda}_{j}}{2}\rangle$ and the spin factor $I_{\text{spin}}=\langle [(cs)^{s_a}(\bar q\bar q)^{s_b}]^S| \hat V_s|[(c\bar q)^{s_c}({s\bar q})^{s_d}]^S$ where $\hat V_s$ is taken as $\mathbf 1$ for the Coulomb and linear confinement interactions, and $\mathbf{s}_{i}\cdot\mathbf{s}_{j}$ for the hyperfine potential. $s_a$ and $s_b$ are the spin of the diquark and antidiquark in the tetraquark state. $s_c$ and $s_d$ are those of the two mesons in the final state. The $3_c\otimes \bar 3_c$ and $\bar 6_c\otimes 6_c$ represent the color configurations $[(\bar c\bar s)_{3_c}(qq)_{\bar 3_c}]_{1_c}$ and {$[(\bar c\bar s)_{\bar 6_c}(qq)_{\bar 6_c}]_{1_c}$}, respectively. }\label{tab:decayfactor}
 \centering
\begin{tabular}{c|ccccc|cccccccc|ccccc}
\toprule[1pt]
 & & & & \multicolumn{2}{c|}{$\langle\frac{\mathbf{\lambda}_{i}}{2}\cdot\frac{\mathbf{\lambda}_{j}}{2}\rangle$} & \multicolumn{8}{c|}{$J^{P}=0^{+}$} & \multicolumn{2}{c}{$J^{P}=2^{+}$} & & & \tabularnewline
\midrule[1pt] $ (s_a,s_b, s_c,s_d)$& & & &
\multirow{1}{*}{$3_c\otimes \bar{3}_{c}$} & \multirow{1}{*}{$\bar
6_{c}\otimes{6}_{c}$} &
\multicolumn{2}{c}{$(1,1,1,1)$} &
\multicolumn{2}{c}{$(1,1,0,0)$} &
\multicolumn{2}{c}{$(0,0,1,1)$} &
\multicolumn{2}{c|}{$(0,0,0,0)$} &
\multicolumn{2}{c}{$(1,1,1,1)$} & & & \tabularnewline \hline
 & & & & & & $\mathbf{1}$ & $\mathbf{s}_{i}\cdot\mathbf{s}_{j}$ & $\mathbf{1}$ & $\mathbf{s}_{i}\cdot\mathbf{s}_{j}$ & $\mathbf{1}$ & $\mathbf{s}_{i}\cdot\mathbf{s}_{j}$ & $\mathbf{1}$ & $\mathbf{s}_{i}\cdot\mathbf{s}_{j}$ & $\mathbf 1$ & $\mathbf{s}_{i}\cdot\mathbf{s}_{j}$ & & & \tabularnewline

$C_1$ & & & & $\frac{2}{9}\sqrt{3}$ & $\frac{\sqrt{6}}{9}$ &
$-\frac{1}{2}$ & $-\frac{1}{8}$ & $-\frac{\sqrt{3}}{2}$ &
$-\frac{\sqrt{3}}{8}$ & $-\frac{\sqrt{3}}{2}$ &
$\frac{3\sqrt{3}}{8}$ & $\frac{1}{2}$ & $-\frac{3}{8}$ & 1 &
$\frac{1}{4}$ & & & \tabularnewline
$C_2$ & & & & $\frac{2}{9}\sqrt{3}$ & $\frac{\sqrt{6}}{9}$ &
$-\frac{1}{2}$ & $-\frac{1}{8}$ & $-\frac{\sqrt{3}}{2}$ &
$-\frac{\sqrt{3}}{8}$ & $-\frac{\sqrt{3}}{2}$ &
$\frac{3\sqrt{3}}{8}$ & $\frac{1}{2}$ & $-\frac{3}{8}$ & 1 &
$\frac{1}{4}$ & & & \tabularnewline
$T_1$ & & & & $-\frac{2}{9}\sqrt{3}$ & $-\frac{\sqrt{6}}{9}$ &
$-\frac{1}{2}$ & $\frac{5}{8}$ & $-\frac{\sqrt{3}}{2}$ &
$\frac{\sqrt{3}}{8}$ & $-\frac{\sqrt{3}}{2}$ & $\frac{\sqrt{3}}{8}$
& $\frac{1}{2}$ & $\frac{3}{8}$ & 1 & $\frac{1}{4}$ & & &
\tabularnewline
$T_2$ & & & & $-\frac{2}{9}\sqrt{3}$ & $-\frac{\sqrt{6}}{9}$ &
$-\frac{1}{2}$ & $\frac{5}{8}$ & $-\frac{\sqrt{3}}{2}$ &
$\frac{\sqrt{3}}{8}$ & $-\frac{\sqrt{3}}{2}$ & $\frac{\sqrt{3}}{8}$
& $\frac{1}{2}$ & $\frac{3}{8}$ & 1 & $\frac{1}{4}$ & & &
\tabularnewline \midrule[1pt]

 & \multicolumn{18}{c}{$J^{P}=1^{+}$}\tabularnewline
\midrule[1pt] $ (s_a,s_b,s_c,s_d)$ &
\multicolumn{2}{c}{$(0,1,1,0)$} &
\multicolumn{2}{c}{$(0,1,0,1)$} &
\multicolumn{2}{c}{{$(0,1,1,1)$}} &
\multicolumn{2}{c}{$(1,0,1,0)$} &
\multicolumn{2}{c}{$(1,0,0,1)$} &
\multicolumn{2}{c}{{$(1,0,1,1)$}} &
\multicolumn{2}{c}{{$(1,1,1,0)$}} &
\multicolumn{2}{c}{$(1,1,0,1)$} &
\multicolumn{2}{c}{$(1,1,1,1)$}\tabularnewline
 \hline
 & $\mathbf{1}$ & $\mathbf{s}_{i}\cdot\mathbf{s}_{j}$ & $\mathbf{1}$ & $\mathbf{s}_{i}\cdot\mathbf{s}_{j}$ & \multicolumn{1}{c}{$\mathbf{1}$} & $\mathbf{s}_{i}\cdot\mathbf{s}_{j}$ & $\mathbf{1}$ & $\mathbf{s}_{i}\cdot\mathbf{s}_{j}$ & $\mathbf{1}$ & $\mathbf{s}_{i}\cdot\mathbf{s}_{j}$ & \multicolumn{1}{c}{$\mathbf{1}$} & $\mathbf{s}_{i}\cdot\mathbf{s}_{j}$ & \multicolumn{1}{c}{$\mathbf{1}$} & $\mathbf{s}_{i}\cdot\mathbf{s}_{j}$ & \multicolumn{1}{c}{$\mathbf{1}$} & $\mathbf{s}_{i}\cdot\mathbf{s}_{j}$ & \multicolumn{1}{c}{$\mathbf{1}$} & $\mathbf{s}_{i}\cdot\mathbf{s}_{j}$\tabularnewline
$C_1$ & $\frac{1}{2}$ & $-\frac{3}{8}$ & $\frac{1}{2}$ & $-\frac{3}{8}$
& \multicolumn{1}{c}{$\frac{1}{\sqrt{2}}$} & $-\frac{3}{4\sqrt{2}}$
& $\frac{1}{2}$ & $\frac{1}{8}$ & $\frac{1}{2}$ & $\frac{1}{8}$ &
$-\frac{1}{\sqrt{2}}$ & $-\frac{1}{4\sqrt{2}}$ &
\multicolumn{1}{c}{$-\frac{1}{\sqrt{2}}$ } & $-\frac{1}{4\sqrt{2}}$
& $\frac{1}{\sqrt{2}}$ & $\frac{1}{4\sqrt{2}}$ & $0$ &
$0$\tabularnewline
$C_2$ & $\frac{1}{2}$ & $\frac{1}{8}$ & $\frac{1}{2}$ & $\frac{1}{8}$ &
\multicolumn{1}{c}{$\frac{1}{\sqrt{2}}$} & $\frac{1}{4\sqrt{2}}$ &
$\frac{1}{2}$ & $-\frac{3}{8}$ & $\frac{1}{2}$ & $-\frac{3}{8}$ &
$-\frac{1}{\sqrt{2}}$ & $\frac{3}{4\sqrt{2}}$ &
\multicolumn{1}{c}{$-\frac{1}{\sqrt{2}}$ } & $-\frac{1}{4\sqrt{2}}$
& $\frac{1}{\sqrt{2}}$ & $\frac{1}{4\sqrt{2}}$ & $0$ &
$0$\tabularnewline
$T_1$ & $\frac{1}{2}$ & $\frac{3}{8}$ & $\frac{1}{2}$ & $-\frac{1}{8}$
& \multicolumn{1}{c}{$\frac{1}{\sqrt{2}}$} & $-\frac{1}{4\sqrt{2}}$
& $\frac{1}{2}$ & $-\frac{1}{8}$ & $\frac{1}{2}$ & $\frac{3}{8}$ &
$-\frac{1}{\sqrt{2}}$ & $\frac{1}{4\sqrt{2}}$
&\multicolumn{1}{c}{$-\frac{1}{\sqrt{2}}$ } & $\frac{1}{4\sqrt{2}}$
& $\frac{1}{\sqrt{2}}$ & $-\frac{1}{4\sqrt{2}}$ & $0$ &
$-\frac{1}{2}$\tabularnewline
$T_2$ & $\frac{1}{2}$ & $-\frac{1}{8}$ & $\frac{1}{2}$ & $\frac{3}{8}$
& \multicolumn{1}{c}{$\frac{1}{\sqrt{2}}$} & $-\frac{1}{4\sqrt{2}}$
& $\frac{1}{2}$ & $\frac{3}{8}$ & $\frac{1}{2}$ & $-\frac{1}{8}$ &
$-\frac{1}{\sqrt{2}}$ & $\frac{1}{4\sqrt{2}}$ &
\multicolumn{1}{c}{$-\frac{1}{\sqrt{2}}$ }& $\frac{1}{4\sqrt{2}}$ &
$\frac{1}{\sqrt{2}}$ & $-\frac{1}{4\sqrt{2}}$ & $0$ &
$\frac{1}{2}$\tabularnewline \bottomrule[1pt]
\end{tabular}
\end{table*}

\end{document}